\documentclass[10pt]{article}
\usepackage{amstex}
\usepackage{times}

\twocolumn

\font\tenhv  = phvb7t at 10pt

\font\elvbf  = cmbx10 scaled 1100

\makeatletter 

\def\@maketitle
   {
   \newpage
   \null
   \vskip .375in 
   \begin{center}
      {\Large \bf \@title \par} 
      \vspace*{24pt} 
      {
      \large 
      \lineskip .5em
      \begin{tabular}[t]{c}
         \@author 
      \end{tabular}
      \par
      } 
      \vskip .5em 
      {
       \large 
      \begin{tabular}[t]{c}
         \@affiliation 
      \end{tabular}
      \par 
      \ifx \@empty \@email
      \else
         \begin{tabular}{r@{~}l}
            E-mail: & {\tt \@email}
         \end{tabular}
         \par
      \fi
      }
      \vspace*{12pt} 
   \end{center}
   } 

\def\abstract
   {%
   \centerline{\large\bf Abstract}%
   \vspace*{12pt}%
   \it%
   }

\def\affiliation#1{\gdef\@affiliation{#1}} \gdef\@affiliation{}

\def\email#1{\gdef\@email{#1}}
\gdef\@email{}

\newlength{\@ctmp}
\newlength{\@figindent}
\long\def\@makecaption#1#2{
   \vskip 10pt
   \setbox\@tempboxa\hbox{\tenhv\noindent #1.~#2} 
   \setlength{\@ctmp}{\hsize}
   \addtolength{\@ctmp}{-\@figindent}\addtolength{\@ctmp}{-\@figindent} 
   \ifdim \wd\@tempboxa >\@ctmp
      \begin{list}{}{\leftmargin\@figindent \rightmargin\leftmargin} 
         \item[]\tenhv #1.~#2\par
      \end{list}
   \else
      \hbox to\hsize{\hfil\box\@tempboxa\hfil} 
   \fi}

\def\section{\@startsection {section}{1}{\z@}
   {14pt plus 2pt minus 2pt}{14pt plus 2pt minus 2pt} {\large\bf}} 
\def\subsection{\@startsection {subsection}{2}{\z@}
   {13pt plus 2pt minus 2pt}{13pt plus 2pt minus 2pt} {\elvbf}}

\makeatother 

\newcommand{\BPP}{\mbox{$\mathcal{BPP}$}}
\newcommand{\QP}{\mbox{$\mathcal{QP}$}}
\newcommand{\EQP}{\mbox{$\mathcal{EQP}$}}
\newcommand{\ZQP}{\mbox{$\mathcal{ZQP}$}}

\newtheorem{theorem}{Theorem}
\newtheorem{lemma}[theorem]{Lemma}
\newtheorem{corollary}[theorem]{Corollary}
\newtheorem{proposition}[theorem]{Proposition}
\newenvironment{proof}{\begin{trivlist}\item[]{\flushleft\bf Proof }}
     {\hspace*{\fill}\raisebox{-1pt}{\boldmath$\Box$}\end{trivlist}}
\newenvironment{proofof}[1]{\begin{trivlist}\item[]{\flushleft\bf Proof
     of #1 }}
     {\hspace*{\fill}\raisebox{-1pt}{\boldmath$\Box$}\end{trivlist}}

\newcommand{\integer}{\Bbb Z}             
\newcommand{\Co}{\Bbb C}                  
\DeclareMathSymbol{\leqslant}{\mathrel}{AMSa}{"36}  
\newcommand{\subgroup}{\leqslant}         
\newcommand\sfrac[1]{\vphantom{\frac1{#1}}{\frac1{\sqrt{\smash[b]{#1}}}}}
\newcommand{\prob}[1]{\mbox{\rm Prob}[#1]}
 
\newcommand{\bra}[1]{\mbox{$\langle #1 |$}}
\newcommand{\ket}[1]{\mbox{$| #1 \rangle$}}
\newcommand{\ketnorm}[1]{\|@,|#1\rangle@,\|} 

\newcommand{\inner}[2]{\mbox{$\langle #1|#2 \rangle$}}

\title{An~Exact Quantum Polynomial-Time Algorithm 
for Simon's Problem}\,%

\author{Gilles Brassard\,%
\thanks{\,Supported in part by Canada's {\sc nserc}, Qu\'ebec's {\sc fcar},
and the Canada Council.}\\[.5em]
\begin{tabular}[t]{c}
D\'epartement IRO\\
Universit\'e de Montr\'eal\\
C.P. 6128, succursale centre-ville\\
Montr\'eal (Qu\'ebec)\\
Canada H3C 3J7\\
{\normalsize Email: {\tt brassard$\mathchar"40$iro.umontreal.ca}}
\end{tabular}
\and
Peter H{\o}yer\,%
\thanks{\,Supported in part by the {\sc esprit} Long Term Research
Programme of the EU under project number 20244 ({\sc alcom-it}).}\\[.5em]
\begin{tabular}[t]{c}
Dept.\ of Math.\ and Computer Science\\ 
Odense University\\
Campusvej~55\\
\mbox{DK--5230} \mbox{Odense~M}\\
Denmark\\
{\normalsize Email: {\tt u2pi$\mathchar"40$imada.ou.dk}}
\end{tabular}
}

\begin{document}
\maketitle

\begin{abstract}
We investigate the power of quantum computers when they are required
to return an answer that is guaranteed to be correct after a time that is
upper-bounded by a polynomial in the worst case.
We show that a natural generalization of Simon's problem can be
solved in this way, whereas previous algorithms required quantum
polynomial time in the expected sense only, without upper bounds on the
worst-case running time.  This is achieved by generalizing both
Simon's and Grover's algorithms and combining them in a novel~way.
It~follows that there is a decision problem that can be solved
in exact quantum polynomial time, which would require expected
exponential time on any classical bounded-error probabilistic computer
if the data is supplied as a black box.
\end{abstract}

\section{Introduction}

According to the modern version of the Church--Turing thesis,
anything that can be computed in polynomial time on a physically
realisable device can be computed in polynomial time on a
probabilistic Turing machine with bounded error probability.
This belief has been seriously challenged by the theory of
quantum computing.
In~particular, \mbox{Simon}~\cite{Simon94} provided the first example of
a problem that can be solved in polynomial time on a
quantum computer, yet any classical bounded-error probabilistic algorithm
would require exponential time if the data is supplied as a black box.
However, Simon's algorithm is polynomial-time in the expected sense:
there is no upper bound on how long it may run on any given
instance if it keeps being unlucky.
(The~same can be said about Shor's celebrated quantum factoring
algorithm~\cite{Shor94}.)

In~this paper, we address
the issue of {\em exact\/} quantum polynomial time,
which concerns problems that quantum computers can
solve in guaranteed worst-case polynomial time with zero
error probability.  Note that this strong \mbox{requirement}
would make randomness useless for classical machines:
anything you can compute on a classical probabilistic
computer with zero error probability in guaranteed worst-case
polynomial time can be done in polynomial time by a {\em deterministic\/}
computer---simply run the probabilistic algorithm with an arbitrarily
fixed sequence of coin ``tosses''.

The study of exact quantum polynomial time is not new.
The very first algorithm ever designed to demonstrate an advantage
of quantum computers over classical computers,
due to Deutsch and Jozsa~\cite{DJ92}, was of this exact nature.
However, it solved a problem that could be handled just as efficiently
with a classical probabilistic computer, provided an arbitrarily
small (one-sided) error probability is tolerated.
Later, Bernstein and Vazirani provided a
relativized problem that can be solved in exact
quantum polynomial time, but not in time $n^{o(\log n)}$ on any classical
bounded-error probabilistic machine~\cite{BV93}.
More recently, we constructed such a problem that
would require exponential time on any classical bounded-error
probabilistic machine~\cite{BH96}.
None of these problems were decision problems.\,%
\footnote{\,\samepage The Deutsch--Jozsa problem gives rise to an oracle
decision problem~\cite{BB92,BB94}.
Also, in the soon-to-be-published journal version of their
paper, \mbox{Bernstein} and Vazirani extend their result to a
decision problem~\cite{BV97}.}
Here we recast Simon's problem in a natural group-theoretic framework,
we generalize it, and we give an exact quantum polynomial-time
\mbox{algorithm} to solve~it.  This provides the first evidence
of an \mbox{exponential} gap between the power
of exact quantum computation and that of classical bounded-error
probabilistic computation, even for decision problems.

Of independent interest are the techniques developed to obtain our results.
Two~of the most fundamental techniques discovered so far in the field of
quantum computation are Simon's~\cite{Simon94} and Grover's~\cite{Grover96}.
Here, we generalize both techniques and we show for the first time that
they can be made to work together toward a common goal:
our algorithm crucially requires the availability of both these tools.

In this paper, we shall use the term \ZQP--algorithm (resp.~\QP--algorithm)
to denote an algorithm that runs in expected (resp.~guaranteed worst-case)
polynomial time on the quantum computer to solve an arbitrary problem.
In~particular, \ZQP{} (resp.~\QP{}) is the class of {\em decision\/}
problems that allow a \ZQP--algorithm (resp.~a \QP--algorithm).\,%
\footnote{\,\samepage\QP{} has been called \EQP{} (``E'' for Exact) by some
authors~\cite{BV93}.}
To~summarize our results, Simon gave a \ZQP{}--algorithm for his problem;
we~generalize it and give a \QP{}--algorithm.  This allows for the
construction of an oracle under which there is a decision problem
in \QP{} that would not only lie outside of the classical class
\BPP{}---which was known already~\cite{BV97}---but that would require
exponential time on any classical bounded-error probabilistic computer.

\section{Simon's subgroup problem}\label{sec:prb}
We~first state Simon's problem~\cite{Simon94}.  Let \mbox{$n \geq 1$} be any
integer and $R$ any set representable on a quantum computer.  Let
$(\oplus) : \{0,1\}^n \times \{0,1\}^n \rightarrow \{0,1\}^n$ denote the
bitwise exclusive-or, written using infix notation.
\begin{description}
\item[Given:] An integer $n \geq 1$ and a function $\rho:\{0,1\}^n
    \rightarrow R$.
\item[Promise:] There exists a nonzero element $s \in \{0,1\}^n$ such
    that for all $g,h \in \{0,1\}^n$, \mbox{$\rho(g)=\rho(h)$} if and
    only if $g=h$ or $g=h \oplus s$.
\item[Problem:] Find~$s$.
\end{description}
We say of such a function $\rho$ that it {\em fulfills Simon's promise\/}
with respect to~$s$.

There is a nice group-theoretic interpretation and generalization for
Simon's problem, and since that interpretation also helps simplify the
notation, we shall use~it.  Hence, we reformulate Simon's problem as
follows.

Let~$\integer_2 = \{0,1\}$ denote the additive group of two elements
with addition denoted by~$\oplus$.  For any given integer $n \! \geq \! 1$,
let~$G$ denote the group $\langle \integer_2^n, \oplus \rangle$.  For
any subset $X \subseteq G$, let $|X|$ denote the cardinality of~$X$ and
let~$\langle X \rangle$ denote the subgroup generated by~$X$.
A~subset~$X$ of a set~$Y$ is {\em proper\/} if $X \neq Y$.  A~subset~$X
\subseteq G$ is {\em linearly independent\/} in~$G$ if no proper subset
of~$X$ generates~$\langle X \rangle$.  If~$H \subgroup G$ is a subgroup
then $g \in G$ is called a {\em representative\/} for the coset $g
\oplus H$.

Define a bilinear map $G \times G \rightarrow \integer_2$~by 
\begin{equation}\label{eq:bilinear}
g \cdot h = \Big(\sum_{i=1}^n g_i h_i\Big)~\text{mod~2}
\end{equation}
where $g = (g_1,\dots,g_n)$ and $h = (h_1,\dots,h_n)$.  For any subgroup
$H \subgroup G$, let
\begin{equation}\label{eq:perpdef}
H^\perp = \{g \in G \,|\, g \cdot h = 0 \text{ for all } h \in H\} 
\end{equation}
denote the {\em orthogonal subgroup} of~$H$.  For any subgroups
\mbox{$K \subgroup H \subgroup G$}, let \mbox{$[H:K]$} denote the index of~$K$
in~$H$, that is, the number of cosets of~$K$ in~$H$.  Note that, for all
subgroups \mbox{$H \subgroup G$}, we have
\mbox{$(H^\perp)^\perp = H$} and \mbox{$|H^\perp| = [G:H]$}.  Using this
terminology, we state the following problem.
\begin{description}
\item[Given:] An integer $n \geq 1$ and a function $\rho:G =
    \integer_2^n \rightarrow R$.
\item[Promise:] There exists a subgroup $H_0 \subgroup G$ such that
    $\rho$ is constant and distinct on each coset of~$H_0$.
\item[Problem:] Find a generating set for~$H_0$.
\end{description}
We say of such a function $\rho$ that it {\em fulfills Simon's promise\/}
with respect to subgroup~$H_0$.

In~Simon's original problem~\cite{Simon94}, $H_0$ is assumed to have
order~2, that is, $H_0 = \{0,s\}$ for some $s \in G$ and the problem is
then to find~$s$.  We~shall, however, in the rest of this paper, refer
to the above problem as {\em Simon's subgroup problem}.  Simon gave
in~\cite{Simon94} a very simple and beautiful quantum algorithm for
solving the subgroup problem.  We~now review the main ideas behind that
algorithm, but we use a
language which is rather different from Simon's.

A~first crucial observation is that, given a generating set for a
subgroup, one can easily (classically or quantumly) \mbox{deduce} a generating
set for its orthogonal subgroup.  This fact is often used in coding
theory: given the generator \mbox{matrix} of a binary linear code,
it allows to compute the generator \mbox{matrix} of its dual.
We~state this formally in
Propositions~\ref{prop:perpperp1} and~\ref{prop:perpperp2} below.

\begin{proposition}\label{prop:perpperp1}
There exists a classical deterministic algorithm that, given a subset
\mbox{$X \subseteq G = \integer_2^n$}, returns a linearly independent subset
of~$G$ that generates the subgroup~\mbox{$\langle X \rangle$}.  Moreover, the
algorithm runs in time polynomial in~$n$ and linear in the cardinality of~$X$.
\end{proposition}

\begin{proposition}\label{prop:perpperp2}
There exists a classical deterministic algorithm that, given a linearly
independent subset \mbox{$X \subset G = \integer_2^n$}, returns a linearly
independent subset of~$G$ that generates the orthogonal subgroup
of~\mbox{$\langle X \rangle$}.  Moreover, the algorithm runs in time
polynomial in~$n$.
\end{proposition}

From these two propositions, and since $(H^\perp)^\perp = H$ for all
subgroups~$H$, it follows that to solve Simon's subgroup problem it
suffices to find a generating set 
\pagebreak
for~$H_0^\perp$.  In~\cite{Simon94},
Simon proved a special case of Theorem~\ref{thm:simon} below, which
gives an efficient quantum algorithm for finding a {\em random\/}
element of~$H_0^\perp$ with respect to the uniform probability
distribution. 

\begin{theorem}\label{thm:simon}
Let $n \geq 1$ be an integer and
\mbox{$\rho: G = \integer_2^n \rightarrow R$} be a function that fulfills
Simon's promise for some subgroup \mbox{$H_0 \subgroup G$}.  Assume that a
quantum algorithm to compute~$\rho$ is given, together with the value of~$n$.

Then there exists a quantum algorithm capable of finding a random element
of the orthogonal subgroup~$H_0^\perp$.  Moreover, the algorithm runs in time
linear in~$n$ and in the time required to compute~$\rho$.
\end{theorem}

We~refer to this algorithm as {\em Simon's subroutine\/} and discuss it
further in the next section.  By~repeating the subroutine until one has
a generating set for~$H_0^\perp$ and then applying Propositions
\ref{prop:perpperp1} and~\ref{prop:perpperp2} above, one has solved
Simon's subgroup problem.  Before this yields an algorithm, however, we
need a procedure to determine {\em when\/} to stop sampling.
Moreover, a bound on the expected number of samples needed to
build a generating set for~$H_0^\perp$ is required
in order to determine the running time of the algorithm.

Consider first the former question of how to determine if a sampled
subset~$Y$ generates~$H_0^\perp$.  If~$H_0$ is of known order~2 (as in
Simon's original paper) then we stop sampling when $Y$ generates a
subgroup of order $|H_0^\perp| = [G:H_0] = 2^{n-1}$.  If~the order of
$H_0$ is unknown then first observe that since $Y \subseteq H_0^\perp$
then $H_0 \subseteq \langle Y \rangle^\perp$ where equality holds if and
only if $Y$ generates~$H_0^\perp$ (and not only a proper subgroup).
In~other words, $Y$ generates $H_0^\perp$ if and only if $\rho$ is
constant on~$\langle Y \rangle^\perp$.  This last condition is easily
checked by first applying Propositions~\ref{prop:perpperp1}
and~\ref{prop:perpperp2} to find a linearly independent set~$X$ that generates
the orthogonal subgroup~\mbox{$\langle Y \rangle^\perp$}, and then evaluating
$\rho$ on~$X$.  It is thus easy to decide when we can stop sampling.

Consider now the latter question of how many times one must repeat
Simon's subroutine in order to obtain a generating set for~$H_0^\perp$.
More generally, given any finite group~$H$, what is the expected number
of elements one must pick from~$H$ in order to have a generating set
for~$H$ when the elements are picked mutually independently with respect
to the uniform probability distribution on~$H$?  There is a simple (easy
to improve) upper bound on this value which can be found as follows.
Let~$K \subgroup H$ be any proper subgroup.  Then the probability that a
randomly picked element in~$H$ is not in~$K$ is at least~$1/2$, so after
an expected number of at most 2 samples, we have picked an element~$z
\in H \setminus K$, and hence $K$ is proper in~$\langle z,K\rangle$.
Since any sequence of proper subgroups in a finite group~$H$ can have
length at most $\log_2|H|$, it follows that after an expected number
of at most~$2 \log_2|H|$ samples we have found a generating set
for~$H$.

By~the above remarks, we can summarize the main steps in Simon's
\ZQP--algorithm for solving his subgroup problem as follows.
Assume we have a quantum polynomial-time algorithm to compute~$\rho$.
By~Theorem~\ref{thm:simon}, we can in polynomial time sample random
elements of the orthogonal subgroup~$H_0^\perp$.  We~have efficient
routines for testing when to stop sampling and for finding $H_0$
from~$H_0^\perp$.  \mbox{Finally}, the expected number of samples needed is
logarithmically bounded in the order of the group, giving an overall
polynomial-time expected running time to find a generating set
for~$H_0$.

In~our approach, we also solve Simon's subgroup problem by first finding
a generating set for~$H_0^\perp$, and we also use the method of finding
repeatedly larger and larger subgroups~$\langle Y \rangle$
of~$H_0^\perp$.  However, instead of finding an element in~$H_0^\perp$
that is not already in~$\langle Y \rangle$ with some bounded
probability, we have discovered a method that {\em guarantees\/} that the
sampled element is taken from the subset $H_0^\perp \setminus \langle Y
\rangle$.  In~addition, our method for finding such an element needs
only time polynomial in~$n$ and in the time required to compute~$\rho$.

\begin{theorem}\label{thm:main}
Let $n \geq 1$ be an integer and
\mbox{$\rho: G = \integer_2^n \rightarrow R$} be a function that fulfills
Simon's promise for some subgroup \mbox{$H_0 \subgroup G$}.  Assume that a
quantum algorithm that computes~$\rho$ without making any measurements
is given, together with the value of~$n$ and a linearly independent subset~$Y$
of the orthogonal subgroup~$H_0^\perp$.

Then there exists a quantum algorithm that returns an \mbox{element}
of~\mbox{$H_0^\perp \setminus \langle Y \rangle$} provided $\,Y$ does not
generate~$H_0^\perp$, and otherwise it returns the zero element.  Moreover,
the algorithm runs in time polynomial in~$n$ and in the time \mbox{required}
to compute~$\rho$.
\end{theorem}

We~postpone the proof of this theorem till Section~\ref{sec:comp}.  Our
new \QP--algorithm for solving Simon's subgroup problem follows easily
from Theorem~\ref{thm:main}.

\begin{theorem}[\QP--algorithm for Simon's problem]
\label{thm:simoninqp}
Let~\mbox{$n \geq 1$} be an integer and
\mbox{$\rho: G = \integer_2^n \rightarrow R$}
be a function that fulfills
Simon's promise for some subgroup \mbox{$H_0 \subgroup G$}.
Then given a
quantum polynomial-time (in~$n$) algorithm to compute~$\rho$
without making measurements, there exists a
\QP--algorithm to find a generating set for~$H_0$.
\end{theorem}

\begin{proof}
The algorithm consists of two stages.  In~the first stage, we find a
generating set for~$H_0^\perp$ as follows.  We~initialize a counter
$i=0$ and set $Y^{(i)} = \emptyset$ to reflect the fact that we initially
do not know any nontrivial elements of the orthogonal subgroup~$H_0^\perp$.

We~then compute the following process.  We~apply Theorem~\ref{thm:main},
giving an element~$z^{(i+1)} \in H_0^\perp$.  If~the outcome $z^{(i+1)}$
is the zero element then we terminate the first stage.  Otherwise, we
set $Y^{(i+1)} = Y^{(i)} \cup \{z^{(i+1)}\}$ and increment the
counter~$i$ by~1.

We~repeat this process until we finally measure the zero element and
then we terminate the first stage.  Note that each of the
subsets~$Y^{(j)}$ $(0 \leq j \leq i)$ is linearly independent by
Theorem~\ref{thm:main}.  Moreover, by the same theorem, since the final
measured element~$z^{(i+1)}$ is the zero element we know that $Y^{(i)}$
generates the orthogonal subgroup.

In~the second stage, we apply Proposition~\ref{prop:perpperp2} on the
set~$Y^{(i)}$ to find a generating set for~$H_0$.  This~completes our
proof of the existence and correctness of the algorithm.

Any linearly independent set in~$G=\integer_2^n$ can have cardinality at
most~$n$ and hence the algorithm applies Theorem~\ref{thm:main} at
most~$n$ times, each taking time polynomial in~$n$.  Since the final
application of Proposition~\ref{prop:perpperp2} also runs in polynomial
time, the claimed running time follows.
\end{proof}

\section{Simon's quantum algorithm}\label{sec:alg}
We~assume in this extended abstract that the reader is familiar with the
basic notions of quantum
computing \mbox{\cite{Barenco96,Berthiaume97,Brassard95}}.  We~denote a
register holding $m$~qubits, all in the zero state, by~\ket{0^m}.  When
its dimension is of no importance, we sometimes just write~\ket{\mathbf
0}.  For any nonempty subset \mbox{$X \subseteq G$}, let \ket{X} denote
the equally-weighted superposition $\sfrac{|X|} \sum_{x \in X}\ket{x}$.
In~particular, if $g \oplus H_0$ is a coset of~$H_0$, then \ket{g \oplus
H_0} denotes the superposition $\sfrac{|H_0|}\sum_{h \in H_0} \ket{g
\oplus h}$.  For any nonempty subset $X \subseteq G$ and any element $g
\in G$, let $\ket{\phi_g X}$ denote the superposition $\sfrac{|X|}
\sum_{x \in X}(-1)^{g \cdot x}\ket{x}$.

Define the one-bit Walsh-Hadamard transform
\[ \mathbf W_2 = \sfrac{2} \sum_{i,j=0}^1 (-1)^{ij} \ket{i}\bra{j} \, . \]
With respect to the ordered basis~\mbox{$(\ket{0},\ket{1})$}, this reads 
\mbox{$\mathbf W_2=\smash{\frac{1}{\sqrt 2}
\left(\begin{smallmatrix}1&\phantom{-}1\\1&-1\end{smallmatrix} \right)}$}. 
Let $\mathbf W_2^n$ denote the Walsh-Hadamard transform applied on each qubit
of a system of $n$~qubits.  The~result of applying
$\mathbf W_2^n$ to \ket{g}, where \mbox{$g \in G$}, is the
\mbox{superposition}~$\ket{\phi_g G}$.
Moreover, for any subgroup \mbox{$K \subgroup G$} and any elements
\mbox{$g,h \in G$},
\begin{equation}
\mathbf W_2^n \ket{\phi_h (g \oplus K)} 
   = (-1)^{g \cdot h} \ket{\phi_g (h \oplus K^\perp)}.
\end{equation}
Thus, by applying the Walsh-Hadamard transform, the subgroup is mapped
to its orthogonal subgroup, and the phases translate to a coset and
vice versa.

A~classical function~$f$ is evaluated reversibly by the
operation~$\mathbf U_{\!f}\!$ which maps $\ket{x}\ket{y}$ to
$\ket{x}\ket{y \oplus f(x)}$~\cite{Bennett73}.  Note that a second
application of~$\mathbf U_{\!f}\!$ will restore the second register to
its original value since $\ket{x}\ket{y \oplus f(x) \oplus f(x)} =
\ket{x}\ket{y}$.

Let~$T_0$ be a transversal of~$H_0$ in~$G$, that is, a subset \mbox{$T_0
\subseteq G$} that consists of exactly one representative from each
coset of~$H_0$.  Simon's subroutine for finding a random \mbox{element}
of~$H_0^\perp$, working on the initial state \ket{0^n}\ket{\mathbf 0},
can be \mbox{described} as follows.

\pagebreak
\noindent
{\sc Simon's subroutine}
\begin{enumerate}
\item Apply the inverse of
      transform~$\mathbf W_2^n$ to the first register\,\footnote{\,
      \samepage Of course, we could apply $\mathbf W_2^n$ rather
      than its inverse since this transformation is self-inverse.
      Nevertheless, it is more natural to think of the \mbox{operation} in
      terms of the inverse of $\mathbf W_2^n$, especially if we wish
      to extend the notion to non-Abelian groups.}
      producing an equally-weighted superposition of all
      elements in the group~$G$, \newline
      \[\sfrac{2^n} \sum_{g \in G} \ket{g}\ket{\mathbf 0}.\]
\item Apply $\mathbf U_\rho$, producing a superposition of all cosets
      of~$H_0$, 
      \[\sfrac{2^n} \sum_{g \in G} \ket{g}\ket{\rho(g)} =
      \sfrac{|T_0|} \sum_{t \in T_0} \ket{t \oplus H_0}\ket{\rho(t)}.\]
\item Apply $\mathbf W_2^n$ to the first register, producing a
      superposition over the orthogonal subgroup~$H_0^\perp$, 
      \begin{equation}\label{eq:aftersimon}
      \ket{\Psi} = \sfrac{|T_0|} \sum_{t \in T_0} \ket{\phi_t
      H_0^\perp}\ket{\rho(t)}.\end{equation}
\end{enumerate}
\medskip
Suppose we measure the first register in the resulting
\mbox{superposition}~$\ket{\Psi}$.  Let~$z$ be the outcome.  It~is immediate
that $z$ is a random element of the orthogonal subgroup~$H_0^\perp$, which
proves Theorem~\ref{thm:simon} above and implies \mbox{Simon's}
\mbox{\ZQP--algorithm} for solving his subgroup problem.

Now, consider the crucial cause in Simon's algorithm why it is not a
\QP--algorithm.  Suppose we have already found an independent set
\mbox{$Y \! \subset \! H_0^\perp$} that generates only a proper subgroup
of~$H_0^\perp$. Then, what we would like is to measure an
element~\mbox{$z \in H$} such that \mbox{$Y \cup \{z\}$} is also linearly
independent.  However, Simon's algorithm promises only that $z$
preserves independence with some probability.  Our~approach to finding $z$
so that \mbox{$Y \cup \{z\}$} is {\em certain\/} to be linearly independent
consists in
solutions to the following two subproblems.  Suppose we have written
$H_0^\perp$ as the internal direct sum of two subgroups,
\mbox{$H_0^\perp = K \oplus \langle Y \rangle$} for some nontrivial
\mbox{$K \subgroup H_0^\perp$}.  Then our solution, informally, consists of
two parts.
\begin{enumerate}
\item We give a method for transforming \ket{H_0^\perp} into \ket{K}.
\item We give a method for transforming \ket{K} into \ket{X} where $X
      \subseteq (K \setminus \{0\})$ is a nonempty subset consisting
      only of some of the nonzero elements of~$K$.
\end{enumerate}

In~the next section, we present our solutions
(Lemma~\ref{lm:smallersg} and Lemma~\ref{lm:AA}, respectively) to
these two problems.  From these, we then prove Theorem~\ref{thm:main}
stated above.  Our new \mbox{\QP--algorithm} for Simon's subgroup problem
(Theorem~\ref{thm:simoninqp}) is an easy corollary to that theorem.

\section{Our new \QP--algorithm}\label{sec:qpalg}
This section consists of three subsections.  The first two contain our
solutions to the two above-mentioned subproblems, while we combine them
in the last subsection to prove Theorem~\ref{thm:main}.

\subsection{Shrinking a subgroup} 
If~we apply Simon's subroutine on the initial zero state
\ket{0^n}\ket{\mathbf 0}, our quantum system is afterwards in
superposition~\ket{\Psi} defined in Equation~\ref{eq:aftersimon}.
In~particular, for every \mbox{element}~$t$ in a transversal for~$H_0$, the
first register holds the superposition \ket{\phi_t H_0^\perp}.  If~we
measure this register then we will measure a random element
of~$H_0^\perp$.  Now, suppose we have earlier measured a nonzero
element~\mbox{$y \in H_0^\perp$}.  Then we would like not to measure $y$
once again, but rather some other element.

The following lemma provides us with a routine that ensures we will not
measure~$y$ again.  Since \mbox{$y=(y_1,\dots,y_n)$} is nonzero, it has some
nonzero \mbox{entry}, say \mbox{$y_j=1$}.  The~idea is to use the
\mbox{$j$th} entry in the first register to change
subgroup $H_0^\perp$ into the smaller subgroup
\mbox{$K=\{(h_1,\dots,h_n) \in H_0^\perp \,|\, h_j=0\}$} of all elements
in $H_0^\perp$ with 0 in that entry.  It~follows then that we shall not
obtain~$y$ again if we measure the first register in this new superposition.

\begin{lemma}\label{lm:shrinkone}
Let \mbox{$H \subgroup G$} be a nontrivial subgroup and let
\mbox{$y = (y_1,\dots,y_n) \in H$} be a known nonzero element.
Let~$j$ be such that~\mbox{$y_j=1$} and let~$K$ denote the
subgroup \mbox{$\{(h_1,\dots,h_n) \in H \,|\, h_j=0\}$}.  

Then there exists a quantum routine that, for all $g \in G$, given
\ket{\phi_g H}\ket{0} returns \ket{\phi_g K}\ket{g \cdot y}.  Moreover,
the routine runs in time linear in~$n$ and it uses no measurements.
\end{lemma}

\begin{proof}
The routine consists of three unitary operations.  Initially, we apply
the controlled {\sc not} operation where the \mbox{$j$th} qubit in the
first register is the control bit and the second register holds the
target bit.  Applying this operation on the input \ket{\phi_g H}\ket{0}
produces
\begin{multline*} 
\sfrac{|H|} \sum_{h \in H} (-1)^{g \cdot h} \ket{h}\ket{h_j} \\
= \sfrac{2} \sum_{i \in \integer_2} (-1)^{g \cdot (iy)} \bigg( \sfrac{|K|}
\sum_{k \in K} (-1)^{g \cdot k} \ket{(iy) \oplus k}\bigg) \ket{i}
\end{multline*}
where \mbox{$h=(h_1,\dots,h_n) \in H$}, \mbox{$0y=0$} and
\mbox{$1y=y$}.  Then, if the second register holds a~1, we apply
the operator defined by $\ket{x} \mapsto \ket{x \oplus y}$ to the first
register.  This produces
\[\frac1{\sqrt 2} \sum_{i \in \integer_2} (-1)^{g \cdot (iy)}
\bigg(\sfrac{|K|} \sum_{k \in K} (-1)^{g \cdot k} \ket{k}\bigg)\ket{i}\]
which also can be written as
\[\ket{\phi_g K} \bigg(\frac1{\sqrt 2} \sum_{i \in \integer_2}
(-1)^{g \cdot (iy)} \ket{i}\bigg).\]
Finally, we apply~$\mathbf W_2$ to the second register, giving the
superposition in the lemma.  The routine uses no measurements and its
running time is clearly linear in~$n$.
\end{proof}

The above lemma can easily be generalized to the case in which we have
already measured not merely one nonzero element of~$H_0^\perp$, but
any linearly independent subset of~$H_0^\perp$.  The solution is then to
apply the above lemma repeatedly for each element in that subset.

\begin{lemma}\label{lm:smallersg}
Let \mbox{$H \subgroup G$} be a nontrivial subgroup and
\mbox{$\{y^{(1)},\dots,y^{(m)}\} \! \subseteq \! H$}
a known linearly independent set
in~$H$.  Then there exist a subgroup \mbox{$K \! \subgroup \! H$} with
\mbox{$H = K \oplus \langle y^{(1)},\dots,y^{(m)} \rangle$} and a quantum
routine that, for all \mbox{$g \in G$},
returns \mbox{\ket{\phi_g K}\ket{g \cdot y^{(1)},\dots,g \cdot y^{(m)}}}
given \mbox{\ket{\phi_g H}\ket{0^m}}.
Moreover, the routine runs in time linear in~$nm$ and it uses no measurements.
\end{lemma}

\subsection{Removing 0 from a subgroup}\label{sec:zero}
Consider first how much we have gained by using the result from the
previous subsection.  Suppose we first apply Simon's subroutine and then
the routine in Lemma~\ref{lm:smallersg}.  Call this combined
routine~${\mathcal A}'$.  Given a linearly independent set
$\{y^{(1)},\dots,y^{(m)}\} \subseteq H_0^\perp$, routine ${\mathcal
A}'$ produces, on~the input \ket{0^n}\ket{\mathbf 0,0^m}, the
superposition
\begin{equation}\label{eq:afterA0}
\ket{\Psi'} = 
\sfrac{|T_0|} \sum_{t \in T_0} \ket{\phi_t K}\ket{\rho(t),
  t \cdot y^{(1)},\dots,t \cdot y^{(m)}}.
\end{equation}
Here $K$ is given as in Lemma~\ref{lm:smallersg}.  Thus, for every~$t$
in a transversal~$T_0$ for~$H_0$, we hold the superposition \ket{\phi_t
K} in the first register.  By Lemma~\ref{lm:smallersg}, if we measure the
first register, we cannot obtain a previously known element
of~$H_0^\perp$.  Neither can we obtain a nonzero element that is a
linear combination of known elements.  Nevertheless, it remains possible
that we obtain the zero element.

In~this subsection, we show how to avoid measuring the zero element.
This solves the second subproblem mentioned at the end of
Section~\ref{sec:alg}.  Our~solution does not build on a
group-theoretical view as in the previous subsection, but \mbox{instead} on a
general view of~${\mathcal A}'$ as a probabilistic quantum \mbox{algorithm}
that succeeds with some bounded probability.

We~say that a state in the superposition \ket{\Psi'} is {\em good\/} if
it contains a nonzero element in the first register.  States that are
not good are said to be {\em bad}.  The success probability
of~${\mathcal A}'$ is the probability that we measure a good state by
measuring the first register of the system.

If~$K=\{0\}$ is the trivial subgroup then the success probability
of~${\mathcal A}'$ is clearly zero.  Otherwise, ${\mathcal A}'$ succeeds
with probability $1-1/k$ where $k=|K|$ is the order of~$K$.  Thus, we
can initially distinguish between these two cases with high probability,
but not with certainty.  Our result in this subsection is a method to
encapsulate ${\mathcal A}'$ in a larger quantum algorithm that succeeds
with certainty provided \mbox{$K\not=\{0\}$} (see~Theorem~\ref{thm:main}
for an example). 
To~obtain this, consider first the more general problem of improving the
success probability of a probabilistic quantum algorithm, formulated as 
follows.

Suppose we are given a quantum algorithm~$\mathcal A$ that on
\mbox{input}~$\ket{\mathbf 0}$ returns some superposition 
$\ket{\Psi} = \sum_{i
\in I} \ket{i}\ket{\psi_i}$ for some finite index set~$I \subset
\integer$.  Suppose also that~$I$ can be written as the disjunct sum of
two sets $A$ and~$B$ where $A$ corresponds to the ``good'' solutions and
$B$ to the ``bad'' solutions, and suppose we are given a quantum
algorithm to compute the characteristic function~$\chi = \chi_A : I
\rightarrow \{0,1\}$ of~$A$.  Let $\ket{A} = \sum_{i \in A}
\ket{i}\ket{\psi_i}$ denote the superposition of good solutions, and
$\ket{B} = \sum_{i \in B} \ket{i}\ket{\psi_i}$ the superposition of bad
solutions.  Write $\ket{\Psi} = \ket{A} + \ket{B}$.  Let $a =
\inner{A}{A}$ denote the probability that we measure a good solution,
and similarly let $b=\inner{B}{B}$.  Note that $\inner{A}{B}=0$ and
hence $a+b=1$.

Using a generalization of the technique in Grover's
\mbox{algorithm}~\cite{Grover96}, we encapsulate~$\mathcal A$ in a larger
quantum algorithm~$\mathcal Q$ such that the probability that~$\mathcal
Q$ returns a good solution is significantly better compared to the
probability that~$\mathcal A$ returns a good solution.
In~\cite{BBHT96}, it is shown that if $\mathcal A = \mathbf W_2^n$ is
the Walsh-Hadamard transform and the probability of success of~$\mathcal
A$ is exactly one quarter (\mbox{$a=1/4$}) then $\mathcal Q$ can be
constructed such that it succeeds with certainty.

For our purpose, we require a similar technique which applies in the
case that $\mathcal A$ is any quantum algorithm that uses no
measurements and has success probability exactly one half (\mbox{$a=1/2$}).
To~obtain this result, we use complex phases, whereas in Grover's
original algorithm only the real phases~$\pm 1$ are
needed~\cite{Grover96}.  Let $\imath = \sqrt{-1}$ denote the square root
of~$-1$.  (Do~not confuse imaginary $\imath$ with integer~$i$.)  The~formal
setting and the lemma are as follows.

\begin{lemma}\label{lm:AA}
Let $\mathcal A$ be a quantum algorithm that uses no measurements and
that given \ket{\mathbf 0} returns \mbox{$\ket{\Psi} = \sum_{i \in I}
\ket{i}\ket{\psi_i}$} for some finite index set~\mbox{$I \subset \integer$}.
Let \mbox{$\chi : I \rightarrow \{0,1\}$} be any Boolean function.
Define
\[ \begin{array}{ll}
A = \{i \in I \,|\, \chi(i)=1\} &
B = \{i \in I \,|\, \chi(i)=0\} \\
\ket{A} = \sum_{i \in A} \ket{i}\ket{\psi_i} &
\ket{B} = \sum_{i \in B} \ket{i}\ket{\psi_i} \\
a = \inner{A}{A} &
b = \inner{B}{B} \, .
\end{array} \]
Then there exists a quantum algorithm~$\mathcal Q$ that on input
\ket{\mathbf 0} returns $k\ket{A} + l\ket{B}$ where $k=2\imath(1-a)-1$
and \mbox{$l=\imath(1-2a)$}.  In~particular, if $a = \frac12$ then the
result is $(\imath-1) \ket{A}$.  If~$a=0$ then \ket{A} has norm zero and
hence the result is $\imath \ket{B}$.  Moreover, $\mathcal Q$ runs in
time linear in the number of qubits and in the times required to compute
$\mathcal A$ and~$\mathbf U_{\chi}$\,, and it uses no measurements.
\end{lemma}

\begin{proof}
First note that \mbox{$B = I \setminus A$} and that
\mbox{$\ket{\Psi} = \ket{A} + \ket{B}$} can be written as a sum of ``good''
and ``bad'' solutions with inner product zero, \mbox{$\inner{A}{B} = 0$}. 
Note also that the probabilities to measure a ``good'' or ``bad'' solution sum
to~1: \mbox{$a+b=1$}.  For every \mbox{$k,l \in \Co$ 
with $|k|^2a + |l|^2b =1$},
define the normalized superposition
\mbox{$\ket{\Psi(k,l)} = k \ket{A} + l \ket{B}$}.  Here $|x|$ denotes the norm
of \mbox{$x \in \Co$}.  Note that
\mbox{$\ket{\Psi(1,1)} = \ket{\Psi} = {\mathcal A} \ket{\mathbf 0}$}.

Now, instead of measuring $\ket{\Psi(1,1)}=\ket{\Psi}$ immediately, we
add one Grover iteration before the measurement.  This Grover iteration
is not the one from Grover's paper~\cite{Grover96} but a generalized
version defined as follows.  Let the phase-change operator~${\mathbf S}_A$
be defined~by
\[\mathbf S_A \ket{i}\ket{\psi_i} = \begin{cases} 
\imath\,\ket{i}\ket{\psi_i} & \text{if $i \in A$}\\
\hphantom{\imath\,}\ket{i}\ket{\psi_i} & \text{if $i \in B$.} \end{cases}\]
In~a similar manner, let $\mathbf S_{\{0\}}$ be the operator that
changes the phase by $\imath$ if and only if the state is the zero state.

Define the {\em Grover iteration\/}
as
\[ {\mathbf G} = {\mathcal A} \,\mathbf S_{\{0\}}\,{\mathcal A}^{-1}
  \,\mathbf S_A.\] 
Straightforward calculations show that applying~$\mathbf G$ on a
\mbox{superposition} of the form \ket{\Psi(k,l)} has the same kind of
\mbox{effect} as the one in~\cite{Grover96}.  In~particular, we have
\[\mathbf G\ket{\Psi(1,1)} 
  = \ket{@,\Psi(2\imath(1-a)-1,\imath(1-2a))}. \]
Let $\mathcal Q$ be the quantum algorithm in which we first apply
${\mathcal A}$ and then~$\mathbf G$.  Then applying~$\mathcal Q$ on
input \ket{\mathbf 0} produces
\begin{align*}
{\mathcal Q} \ket{\mathbf 0} &= \mathbf G \ket{\Psi(1,1)} \\
 &= (2\imath(1-a)-1)\ket{A} + \imath(1-2a)\ket{B},
\end{align*}
and the first part of the lemma follows.

The phase-change operator $\mathbf S_{\{0\}}$ can be applied in
time linear in the number of qubits, while~$\mathbf S_A$ can be
applied by computing~$\mathbf U_\chi$ twice and doing a constant amount
of additional work~\cite{BBC+95}.  Hence, $\mathcal Q$ runs in time linear
in the number of qubits and in the times required to compute $\mathcal A$
and~$\mathbf U_{\chi}$. \looseness=-1
\end{proof}

\subsection{Composing our new \QP--algorithm}\label{sec:comp}
By~Lemma~\ref{lm:AA}, we can take a quantum algorithm~$\mathcal A$ and
construct a new quantum algorithm~$\mathcal Q$ such that
if~$\mathcal A$ succeeds with probability zero then so does~$\mathcal
Q$, and if~$\mathcal A$ has success probability 1/2 then~$\mathcal Q$
succeeds with certainty.  Consider the algorithm~${\mathcal A}'$
defined in the beginning of Subsection~\ref{sec:zero}.  It~succeeds with
probability zero if $K=\{0\}$ is trivial, and otherwise with probability
$1-1/k$ where $k=|K|$ is the order of~$K$.

At~first glance, it seems that we cannot apply Lemma~\ref{lm:AA} since
${\mathcal A}'$ succeeds with too large a probability.  Fortunately, we
can get around this problem by redefining what we mean by a state in the
superposition \ket{\Psi'} given in~(\ref{eq:afterA0}) being good.  Fix
an~$i$ with $1 \leq i \leq n$.  Redefine a state to be good if the
\mbox{$i$th} entry in the first register is~1.  What is now the
success probability~$p_i$ of~${\mathcal A}'$?  If all elements in~$K$
have 0 in the \mbox{$i$th} entry then $p_i=0$.  Otherwise, exactly
half the \mbox{elements} in~$K$ have 0 in the \mbox{$i$th} entry and
exactly half of them have~1, and therefore $p_i=1/2$.  The
success probability~$p_i$ of~${\mathcal A}'$ is thus either zero or
one~half.

Consider the set of probabilities $\{p_1,\dots,p_n\}$.  It~contains one
value for each entry in the first register.  If~$K=\{0\}$ is trivial
then $p_i=0$ for all $1 \leq i \leq n$.  Otherwise, if $K$ is
nontrivial then at least for one $i$ with $1 \leq i \leq n$ we have
$p_i=1/2$.  This suggests that if we apply Lemma~\ref{lm:AA}
once for each of the $n$ different values of~$i$ in order to
improve the success probability of ${\mathcal A}'$
from~$p_i$ to~$2 p_i$, then at least one of the measured elements is a nonzero
element of~$K$ if and only if~$K$ is nontrivial.  We~prove now that this
is indeed the case by giving our proof of Theorem~\ref{thm:main} stated in
Section~\ref{sec:prb}.

\begin{proofof}{Theorem~\ref{thm:main}}
Write $H_0^\perp = K \oplus \langle Y \rangle$ as the direct sum of a
subgroup~$K$ and the subgroup generated by the known elements~$Y \subset
H_0^\perp$.

Let $P = \{ i : 1 \leq i \leq n\}$.  For every $i \in P$, we construct a
quantum algorithm~${\mathcal Q}_i$ that on input
\ket{0^n}\ket{\mathbf 0,0^m} returns an element~$z^{(i)}$ of~$K$ after a
measurement.  We~then construct a larger algorithm which consists of all
the $n$~smaller~${\mathcal Q}_i$ and we show that at least one of the
measured elements~$z^{(i)}$ is nonzero if and only if $K$ is nontrivial.

Fix an $i \in P$.  Define the function $\chi_i : G \rightarrow \{0,1\}$
by $\chi_i(g)=g_i$ where $g=(g_1,\dots,g_n)$.  Thus, $\chi_i(g)$ is~1
if and only if the $i$th entry of~$g$ is~1.

The output of the computation ${\mathcal A}' \ket{0^n}\ket{{\mathbf
0},0^m}$ is the \mbox{superposition} \ket{\Psi'} given in~(\ref{eq:afterA0}).
If~we measure \ket{\Psi'}, let $p_i$ denote the probability that we
obtain a state \ket{g}\ket{x} with the \mbox{$i$th} entry
of~$g$ equal to~1, \mbox{$g_i=1$}.  If~all elements in~$K$ hold a 0 in
that entry then~$p_i$ is zero.  Otherwise, half the \mbox{elements} in~$K$
hold a 1 in that entry and thus, independently of the content of the
second register of~\ket{\Psi'}, we have that \mbox{$p_i=1/2$}.

Suppose we apply Lemma~\ref{lm:AA} on the function~$\chi_i$ and the
algorithm~${\mathcal A}'$ defined above.  Let~${\mathcal Q}_i$ denote
the resulting quantum algorithm.  Consider the content of the first
register in the final superposition.  By Lemma~\ref{lm:AA}, that
register contains only elements from \mbox{$K \subgroup H_0^\perp$},
as did~\ket{\Psi'} originally.  Moreover, each of
these elements holds a 1 in the \mbox{$i$th} entry if and only if
$p_i=1/2$.

Suppose we measure the first register.  Let $z^{(i)} \in K$ be
the outcome.  If $p_i=1/2$ then $z^{(i)}$ is nonzero with certainty.
Otherwise, that is if $p_i=0$, then $z^{(i)}$ may or may not be nonzero.

Consider that we run quantum algorithm ${\mathcal Q}_i$
sequentially for each \mbox{$i \in P$}, and follow each run by a measurement of
the first register.  Suppose $K$ is nontrivial.  Let $g \in K$ be any nonzero
element and let $i_0 \in P$ be so that $g_{i_0}=1$ where
\mbox{$g=(g_1,\dots,g_n)$}.  Then $p_{i_0} = 1/2$ and therefore, with
certainty, the measured element $z^{(i_0)} \in K$ is nonzero.
Now, suppose $K$ is trivial.  Then, for all $i \in P$, we have that
$z^{(i)}$ is the zero element.  This completes the first part of the
\mbox{theorem}.

Consider the overall running time of this composed quantum algorithm.
Each of the transforms~$\mathbf U_{\chi_i}$ can clearly be implemented
in constant time and thus, by Lemma~\ref{lm:AA}, ${\mathcal Q}_i$~runs
in time linear in~$n$ and in the time required to compute~$\mathcal A$.
Since the composed algorithm consists of running each of the $n$
algorithms~${\mathcal Q}_i$ one after the other, it runs in time polynomial
in~$n$ and in the time required to compute~$\mathcal A$ as~well.
\end{proofof}

Having proved Theorem~\ref{thm:main}, we have completed the \mbox{description}
and proof of our new \QP--algorithm for solving Simon's subgroup
problem.

We~end this section with a supplementary remark on the total number of
times we need to compute function~$\rho$.  By~the proof of
Theorem~\ref{thm:simoninqp}, we apply Theorem~\ref{thm:main} at most~$n$
times.  In~each of these applications, we run each of the $n$ quantum
algorithms~${\mathcal Q}_i$ $(i \in P)$ defined above.  Since each
${\mathcal Q}_i$ computes the function~$\rho$ a constant number of times
this gives an upper bound of $O(n^2)$ evaluations of~$\rho$.

We~will show that just $O(n)$ evaluations of~$\rho$ suffices.
First, restate the above counting argument as follows.  For each $i \in
P$, our \QP--algorithm applies ${\mathcal Q}_i$ at most~$n$ times.
Since~$P$ has cardinality~$n$ the number of evaluations of~$\rho$
is~$O(n^2)$.  We~now show that it is suffices to run each
${\mathcal Q}_i$ at most once.

Consider the first time we run~${\mathcal Q}_i$ for $i \in P$.  There
are two cases depending on the outcome~$z^{(i)}$.  If~$z^{(i)}$ is the
zero element then we know that all elements in the subgroup~$K$ (defined
in the beginning of the proof of Theorem~\ref{thm:main}) hold a 0 in
the \mbox{$i$th} entry.  Moreover, this will also be the case in
successive iterations when $K$ has shrunk further.  Therefore,
it would be pointless to run~${\mathcal Q}_i$ again.

On~the other hand, if $z^{(i)}$ is nonzero then it fulfills the
\mbox{requirements} in Theorem~\ref{thm:main} of being a nonzero element
\mbox{preserving} independence.  Thus, we do not need to run any of the
remaining~${\mathcal Q}_i$ algorithms in that application of
Theorem~\ref{thm:main}.  Moreover, since the \mbox{$i$th} entry
in~$z^{(i)}$ is a~1, we can construct our new subgroup~$K$ in the next
applications of Theorem~\ref{thm:main} such that all elements in the new
subgroup~$K$ hold a 0 in the \mbox{$i$th} entry.  This is done by
letting that entry be the control bit in Lemma~\ref{lm:shrinkone}, that
is, by choosing $j=i$.  Thus, also in this case, we do not need to
run~${\mathcal Q}_i$ again.
\looseness=1

\section{Decision problems}
Until now, we have dealt with a version of Simon's problem
that consists of finding a generating set for some subgroup
\mbox{$H_0 \subgroup G = \integer_2^n$}.  
In~Simon's original setting, this subgroup is of order~2
and the problem reduces to finding its unique nonzero element,
called $s$ at the beginning of Section~\ref{sec:prb}.
Recall that we say of such a function that it {\em fulfills Simon's promise\/}
with respect to~$s$.
A~natural question is whether there exists a {\em decision\/} problem
in~\QP{} that would require exponential time to decide on
any classical bounded-error probabilistic computer.
In~this section, we give a positive answer in an appropriate
oracle setting.

This can be achieved in several ways.  The simplest is to note
that our algorithm can distinguish with certainty, after guaranteed
worst-case polynomial time, between a function
\mbox{$\rho:\{0,1\}^n \rightarrow \{0,1\}^n$} that is a bijection
and one that fulfills Simon's promise:
Just apply our general algorithm and see if it turns out zero
or one generator for~$H_0$.  (Recall that Simon's original algorithm
could distinguish these two cases with certainty after {\em expected\/}
polynomial time, or alternatively it could distinguish them
{\em with bounded error probability\/} after guaranteed worst-case polynomial
time.) In~his paper~\cite{Simon94}, Simon proves the existence of an oracle
$\mathcal O$ and a decision problem $L$ such that
(1)~no~classical probabilistic oracle machine that queries $\mathcal O$
fewer than $2^{n/4}$ times on input $1^n$ can decide $L$ with bounded error
probability, and (2)~deciding $L$ given $\mathcal O$ reduces efficiently
and deterministically to the problem of distinguishing between the two types
of functions mentioned above.  It~follows from our algorithm that $L$ can be
decided with certainty in guaranteed worst-case polynomial time on a
quantum computer, given $\mathcal O$ as oracle: 
\mbox{$L \in \QP^{\mathcal O}$}.

Another approach to transforming Simon's problem into a decision problem,
which we find more elegant, is to consider an arbitrary function
\mbox{$\gamma : \{0,1\}^{+} \rightarrow \{0,1\}$},
which is {\em balanced\/} in the sense that there are exactly
$2^{n-1}$ strings \mbox{$x \in \{0,1\}^n$} such that \mbox{$\gamma(x)=b$}
for each \mbox{$b \in \{0,1\}$} and \mbox{$n \ge 1$}.  For~example,
$\gamma(x)$ could be simply the most significant or the least significant bit
of~$x$, or it could be the exclusive-or of all the bits in~$x$.
Consider now an integer $n$ and a function
\mbox{$\rho:\{0,1\}^n \rightarrow \{0,1\}^{n-1}$}
chosen randomly according to the uniform distribution among all
functions that fulfill Simon's promise with respect to
some nonzero \mbox{$s \in \{0,1\}^n$}.
We~prove below (Theorem~\ref{mainthm}) that no subexponential-time classical
probabilistic algorithm can guess $\gamma(s)$ essentially better than at
random, except with exponentially small probability, when $\rho$ is provided as
an oracle. The~probabilities are taken among all choices for~$\rho$,
as~well as the probabilistic choices made by the algorithm,
but {\em not\/} over the possible choice for~$\gamma$:
any fixed balanced $\gamma$ will~do.
It~follows (Corollary~\ref{corol}) that there is an oracle that
simultaneously defeats every classical
algorithm.

\begin{theorem}\label{mainthm}
Fix an arbitrary balanced function~$\gamma$ and an
integer~\mbox{$n \geq 4$}.  Consider an arbitrary classical
probabilistic algorithm that has access to a function oracle
\mbox{$\rho:\{0,1\}^n \rightarrow \{0,1\}^{n-1}$} chosen at random according
to the uniform distribution among all functions that fulfill Simon's
promise with respect to some~$s$.
Assume the algorithm queries its oracle no more than $2^{n/3}$ times.
Then there exists an event $\mathcal E$ such that
(1)~\mbox{$\prob{{\mathcal E}} < 2^{-n/3}$}, and
(2)~If~$\mathcal E$ does {\em not\/} occur then the probability that
the algorithm correctly returns $\gamma(s)$
is less than \mbox{$\frac{1}{2}+2^{-n/3}$}.
The~probabilities are taken over all possible choices of function $\rho$
and the probabilistic choices made by the algorithm.
It~follows that the algorithm cannot guess the value of $\gamma(s)$
with a probability better than \mbox{$\frac{1}{2}+2 \times  2^{-n/3}$}.
\end{theorem}

\begin{proof}
Assume that the algorithm has queried its
oracle on inputs $x_1$, \mbox{$x_2$,\ldots,} $x_k$ for
$x_i \in \{0,1\}^n$, $1 \le i \le k \le 2^{n/3}$.
Without loss of generality, assume that all the queries are distinct.
Let $y_1$, $y_2$,\ldots, $y_k$ be the answers obtained from the
oracle, i.e.~$y_i=\rho(x_i)$ for each~$i$.  Define the event $\mathcal E$
as {\em occurring\/} if there exist $i$ and $j$,
$1 \le i < j \le k$, such that $y_i=y_j$.
Clearly, the algorithm has discovered the secret $s$ when $\mathcal E$
occurs since in that case $s=x_i \oplus x_j$.  This allows the
algorithm to determine $\gamma(s)$ with certainty.
We~have to prove that $\mathcal E$ is very unlikely and that,
unless $\mathcal E$ occurs, the algorithm has so little information
that it cannot guess $\gamma(s)$ significantly
better than at random.

Let \mbox{$X=\{x_1,x_2,\ldots,x_k\}$} be the set of queries to the oracle
and let \mbox{$Y=\{y_1,y_2,\ldots,y_k\}$} be the corresponding answers.
Let \mbox{$W=\{x_i \oplus x_j \,|\, 1 \le i < j \le k\}$},
\mbox{$S=\{0,1\}^n \setminus (W \cup \{0^n\})$} and let
\mbox{$m < k^2$} be the cardinality of~$W$.
Note that $\cal E$ occurs if and only if \mbox{$s \in W$} since
\mbox{$y_i=y_j$} if and only if \mbox{$x_i \oplus x_j=s$}.
If~$\cal E$ does not occur, we say that any \mbox{${\hat s} \in S$}
is {\em compatible\/} with the available data \mbox{because} it is not
ruled out as a possible value for the actual unknown~$s$.
Similarly, given any compatible $\hat s$, we say that
a function \mbox{${\hat \rho}:\{0,1\}^n \rightarrow \{0,1\}^{n-1}$}
is {\em compatible\/} with the available data (and with \mbox{$s=\hat s$})
if \mbox{${\hat \rho}(x_i)=y_i$} for all~\mbox{$i \le k$}, and if
\mbox{${\hat \rho}(x)={\hat \rho}(x')$} if and only if
\mbox{$x \oplus x' = {\hat s}$} for all distinct $x$ and $x'$ in
\mbox{$\{0,1\}^n$}.  

Assume for the moment that $\mathcal E$ has not occurred.
Now we prove that there are exactly \mbox{$(2^n-m-1)((2^{n-1}-k)!)$}
functions that are compatible with the available data.
For~each compatible $\hat s$, exactly
\mbox{$(2^{n-1}-k)!$} of those functions are also compatible with~$s=\hat s$.
It~follows that all compatible values for
$s$ are equally likely to be correct given the available data,
and therefore the only information available about $s$
is that it belongs to~$S$.
For this, consider an arbitrary compatible~$\hat s$.
Define $X'=\{x \oplus {\hat s} \,|\, x \in X\}$.
It~follows from the compatibility of $\hat s$ that $X \cap X'=\emptyset$.
Let \mbox{$Z=\{0,1\}^n \setminus (X \cup X')$}.
Note that \mbox{$x\in Z$} if and only if \mbox{$x \oplus \hat{s} \in Z$}.
Partition $Z$
in an arbitrary way into \mbox{$Z_1 \cup Z_2$} so that \mbox{$x \in Z_1$}
if and only if \mbox{$x \oplus \hat{s} \in Z_2$}.  
The~cardinalities of $Z_1$ and $Z_2$ are \mbox{$(2^n-2k)/2 = 2^{n-1}-k$}.
Now let \mbox{$Y' = \{0,1\}^{n-1} \setminus Y$}, also a set of cardinality
\mbox{$2^{n-1}-k$}. To~each bijection \mbox{$\xi:Z_1 \rightarrow Y'$} there
corresponds a function~$\hat \rho$ compatible with the available data and
\mbox{$s=\hat s$} defined~by
\[ {\hat \rho}(x) ~=~ \left\{
\begin{array}{ll}
y_i & \mbox{if } x=x_i \mbox{ for some } 1 \le i \le k \\
y_i & \mbox{if } x=x_i \oplus {\hat s} \mbox{ for some } 1 \le i \le k \\
\xi(x) & \mbox{if } x \in Z_1 \\
\xi(x\oplus {\hat s}) & \mbox{if } x \in Z_2 \, .
\end{array}
\right. 
\]
The conclusion about the number of compatible functions
follows from the facts that there are
\mbox{$(2^{n-1}-k)!$} such bijections, each possible function
compatible with the available data and $s=\hat s$ is
counted exactly once by this process, and there are
$2^n-m-1$ compatible choices for $\hat s$, each yielding
a disjoint set of functions compatible with the available data.

Still considering the case that $\mathcal E$ has not occurred,
let~\mbox{$A=\{z \in S \,|\, \gamma(z)=1 \}$}
and \mbox{$B=\{z \in S \,|\, \gamma(z)=0 \}$}.
Because we have just seen that each elements of $S$ is equally likely
to be the correct value for~$s$, 
the algorithm's best strategy is to return $1$ if
\mbox{$|A|>|B|$} and $0$ otherwise.
In~the best case (for the algorithm), there are $2^{n-1}$
strings in $A$ and the remaining $2^{n-1}-1-m$ strings are in~$B$,
in which case the guess is correct with probability
\[ \frac{2^{n-1}}{2^n-1-m} \le
   \frac{2^{n-1}}{2^n-k^2} \le
   \frac{2^{n-1}}{2^n-2^{2n/3}} 
   < \frac{1}{2} + 2^{-n/3}
\]
provided $n \ge 4$.

It~remains to prove that the probability that event $\mathcal E$
occurs is exponentially small.  For this, note that all nonzero values
for $s$ are equally likely {\em a~priori\/} and event $\mathcal E$
occurs if and only if \mbox{$s \in W$}.  It~follows that
\[ \prob{{\mathcal E}} = m/(2^n-1) < k^2/2^n \le 2^{-n/3} , \]
where $m$ is the cardinality of~$W$ and $k \le 2^{n/3}$
is the number of oracle queries.

In~conclusion, the probability that the algorithm guesses $\gamma(s)$
correctly is less than
\[ \begin{array}{l}
\prob{{\mathcal E}}+
(1-\prob{{\mathcal E}})\left(\frac{1}{2}+2^{-n/3}\right)\\[2ex]
< 2^{-n/3} + \left(\frac{1}{2} + 2^{-n/3}\right)
= \frac{1}{2} + 2 \times 2^{-n/3} \, . \end{array}\]
\end{proof}

The theorem we have just proven says that no classical probabilistic
algorithm can guess $\gamma(s)$ much better than at random without
spending exponential time on a random function that fulfills Simon's
promise, provided that function is supplied as a black box chosen {\em after\/}
the algorithm has been fixed.  Can~we find a single function that
simultaneously defeats {\em all\/} classical probabilistic algorithms?
The~answer is obviously negative for any fixed finite function.
Nevertheless, the following corollary shows that it is possible to
encode an infinite number of such functions into a single oracle
so that every classical probabilistic algorithm is defeated
infinitely many times.   This exhibits an exponential gap between the power
of exact quantum computation and that of classical bounded-error
probabilistic computation, even for decision problems.

\begin{corollary}\label{corol}
There exists an oracle $\mathcal O$ relative to which there is a decision
problem \mbox{$L \in \QP^{\mathcal O}$} so that, 
for any classical probabilistic
algorithm whose running time is bounded by $2^{n/3}$ on all inputs of size~$n$,
there are infinitely many \mbox{inputs} about which the algorithm decides
membership in $L$ with probability no better than
\mbox{$\frac{1}{2}+2 \times 2^{-n/3}$}.
\end{corollary}

\begin{proof}
Fix some polynomial-time computable balanced function~$\gamma$ 
once and for~all.
For any fixed classical probabilistic algorithm, integer~$n$,
and function \mbox{$\rho:\{0,1\}^n \rightarrow \{0,1\}^{n-1}$}
that fulfills Simon's promise with respect to some~$s$, we say that
the algorithm is {\em defeated\/} by $\rho$ if it cannot guess $\gamma(s)$
with probability better than \mbox{$\frac{1}{2}+2 \times 2^{-n/3}$}
after taking less than $2^{n/3}$ steps.  It~follows directly
from Theorem~\ref{mainthm} that every classical probabilistic
algorithm is defeated by at least one $\rho$
for each value of~\mbox{$n \ge 4$}.
This remains true even if the algorithm is supplied with another arbitrary
fixed oracle, in addition to the oracle for~$\rho$.

Define function $e$ by \mbox{$e(1)=2$} and \mbox{$e(i+1) = 2^{e(i)}$}
for~\mbox{$i \ge 1$}.
Let $\eta$ be an arbitrary function that maps integers to classical
probabilistic algorithms such that every algorithm appears
infinitely many times in the image of~$\eta$.
For~any integer $n$ and functions
\mbox{$\rho:\{0,1\}^n \rightarrow \{0,1\}^{n-1}$}
and \mbox{$\sigma:\{0,1\}^{+} \rightarrow \{0,1\}^{\star}$},
let \mbox{$[\rho,\sigma]:\{0,1\}^{+} \rightarrow \{0,1\}^{\star}$}
denote the function that sends $x$ to $\rho(x)$ for all
\mbox{$x \in \{0,1\}^n$} and to $\sigma(x)$ otherwise.
We~build the required function oracle
\mbox{${\mathcal O}:\{0,1\}^{+} \rightarrow \{0,1\}^{\star}$} by stages:
\mbox{${\mathcal O}(x) = {\mathcal O}_i(x)$} for all
\mbox{$x \in \{0,1\}^n$} such that \mbox{$n < e(i+1)$} and \mbox{$i \ge 1$}.
Initially, ${\mathcal O}_1(x)$ is the string of size \mbox{$n-1$} obtained
by removing the most significant bit of $x$ for each
\mbox{$x \in \{0,1\}^n$} and each \mbox{$nÊ\ge 1$}.
For~each \mbox{$i \ge 2$}, let \mbox{$n=e(i)$} and
\mbox{define} ${\mathcal O}_i$ given ${\mathcal O}_{i-1}$
as follows: choose a function
\mbox{$\rho_i:\{0,1\}^n \rightarrow \{0,1\}^{n-1}$}
that fulfills Simon's promise with respect to some $s_i$ 
in a way that defeats algorithm $\eta(i)$ given
\mbox{$[\rho_i,{\mathcal O}_{i-1}]$} as oracle;
let \mbox{${\mathcal O}_i=[\rho_i,{\mathcal O}_{i-1}]$}.
Finally define
\[ L = \{ 1^{e(i)} \,|\, i \ge 2 \mbox{ and } \gamma(s_i)=1 \} . \]

Note that for each positive integer~$n$, the restriction
of $\mathcal O$ to \mbox{$\{0,1\}^n$} is a function that fulfills Simon's
promise, and therefore \mbox{$L \in \QP^{\mathcal O}$} by the algorithm
given in this paper.  On~the other hand, consider an arbitrary
classical probabilistic algorithm {\sf A} and let $i$ be one of the
infinitely many integers such that \mbox{$\eta(i)=\mbox{\sf A}$}.
\pagebreak
We~know by construction that~$\rho_i$ defeats {\sf A} given
oracle~${\mathcal O}_i$.  This means that {\sf A} cannot guess
$\gamma(s_i)$ significantly better than at random after taking
less than exponentially many steps on input~$1^{e(i)}$.
But~{\sf A} would require exponential time even to {\em formulate\/}
a question of size \mbox{$e(i+1) = 2^{e(i)}$} or bigger for its oracle.
Since \mbox{${\mathcal O}(x)={\mathcal O}_i(x)$} for all
$x$ of size shorter than \mbox{$e(i+1)$}, it follows that
{\sf A} \mbox{behaves} in the same way on input $1^{e(i)}$ whether
it is given ${\mathcal O}$ or ${\mathcal O}_i$ as oracle,
unless it takes exponential time.  Therefore $\rho_i$ defeats {\sf A} given
oracle~${\mathcal O}$ as well.  This happens infinitely \mbox{often} for
each classical probabilistic \mbox{algorithm}, which proves the desired result.
\end{proof}

\section{Other Abelian groups}\label{sec:extensions}
So~far, we have restricted our attention to the Abelian
group~$G=\integer_2^n$.  In~this section, we discuss how these results
generalize to other Abelian groups.  We~start by considering the natural
extension of Simon's problem and subroutine to an arbitrary finite
additive Abelian group.  Our~presentation is kept in group-theoretical
terms and our main tools are three quantum operators defined on the
group.  We~also discuss how our own algorithm generalizes.

For every \mbox{$m \! \geq \! 1$}, let $\integer_m$ denote the additive cyclic
group of order~$m$.  For any given $n$--tuple of positive integers
$(m_1,\dots,m_n)$, let $G = \langle G, +\rangle$ denote the finite
additive Abelian group $\integer_{m_1} \oplus \dots \oplus
\integer_{m_n}$.  We~define the {\em Abelian subgroup problem\/} as follows:
Given group~$G$ and a function~$\rho$ defined on~$G$ and promised to
be constant and distinct on each coset of some unknown subgroup~$H_0
\subgroup G$, find a generating set for~$H_0$.

Our first task is to generalize the concept of orthogonality given by
Equations~\ref{eq:bilinear} and~\ref{eq:perpdef}.  For every \mbox{$m \geq 1$},
let $\omega_m = \exp(2 \pi \imath /m)$ denote the \mbox{$m$th} principal
root of unity.  Let~$\Co^*$ denote the multiplicative group of the
nonzero complex numbers.  Define a bilinear map $\mu : G \times G
\rightarrow \Co^*$~by
\begin{equation}
\mu(g,h) = \sum_{i=1}^n \omega_{m_i}^{g_i h_i}
\end{equation}
where \mbox{$g = (g_1,\dots,g_n)$} and \mbox{$h = (h_1,\dots,h_n)$}.
We~say that an element~\mbox{$g \in G$} is {\em orthogonal\/} to a
subset~\mbox{$X \subseteq G$} if, for all~\mbox{$x \in X$}, we have that
$\mu(g,x)$ is the identity of the group~$\Co^*$, that is, if
\mbox{$\mu(g,x)=1$}.  Note the correspondence to the
bilinear map in Section~\ref{sec:alg}: There, the image was an additive
group with identity~0, while here, the image is a multiplicative group
with identity~1.  For any subgroup \mbox{$H \subgroup G$},~let
\begin{equation}
H^\perp = \{g \in G \,|\, \mu(g,h) = 1 \text{ for all } h \in H\}
\end{equation}
denote the set of all elements in~$G$ orthogonal to~$H$.  Clearly,
$H^\perp$ is a subgroup and we refer to it as the {\em orthogonal
subgroup\/} of~$H$.

For~any subgroups $K \subgroup H \subgroup G$, let $[H:K]$ denote the index
of~$K$ in~$H$.  As~in the simple case $G=\integer_2^n$, we have the
duality relations
\begin{align*}
|H^\perp| &= [G:H]\\
H^{\perp\perp} &= H
\end{align*}
for all subgroups $H \subgroup G$.

We~now define three fundamental quantum operators for the group~$G$.
Together, they extend the ideas and the notation used in
Section~\ref{sec:alg}.  They are the {\em quantum Fourier
transform\/}~${\mathbf F}_G$, the {\em translation operator\/}~${\mathbf
\tau}_t$ $(t \in G)$, and the {\em phase-change operator\/}~${\mathbf
\phi}_h$ $(h \in G)$, defined as follows.
\begin{align*}
{\mathbf F}_G &= \sfrac{|G|} \sum_{g,h \in G} \mu(g,h) \ket{g}\bra{h}\\ 
{\mathbf \tau}_t &= \sum_{g \in G} \ket{t + g}\bra{g}\\
{\mathbf \phi}_h &= \sum_{g \in G} \mu(h,g) \ket{g}\bra{g}
\end{align*}

One~may readily check that these three $G$--operators are unitary.
Note that when \mbox{$G= \integer_2^n$} then the transform ${\mathbf F}_G$ is
just the Walsh-Hadamard transform ${\mathbf W}_2^n$ used in
Section~\ref{sec:alg}.  Unsurprisingly, the Fourier transform maps a
subgroup \mbox{$H \subgroup G$} to its orthogonal subgroup~$H^\perp$,
\[{\mathbf F}_G \ket{H} = \ket{H^\perp}.\]
Moreover, the $G$--operators satisfy the following commutative
laws which we state without proof.
\begin{theorem}[Commutative laws of the {\boldmath $G$}--operators]
\label{thm:comm}
For every $h,t \in G$ we have
\begin{align*}
\mu(h,t) \,\tau_t \, \phi_h &= \phi_h \, \tau_t\\
\mathbf F_G \, \phi_h &= \tau_{-h} \, \mathbf F_G\\
\mathbf F_G \, \tau_t &= \phi_t \, \mathbf F_G \, .
\end{align*}
\end{theorem}

With this setup, we can give a natural extension of \mbox{Simon}'s subroutine,
denoted~${\mathbf U}_G$, for the general Abelian subgroup problem.
\begin{equation}\label{eq:simongeneral}
  {\mathbf U}_G = 
  ({\mathbf F}_G \otimes {\mathbf I}) \circ {\mathbf U}_\rho
  \circ ({\mathbf F}_G^{-1} \otimes {\mathbf I})
\end{equation}
Here, $\mathbf I$ denotes the identity operator, and the notation
\mbox{${\mathbf U}_1 \otimes {\mathbf U}_2$} means applying the unitary
operator ${\mathbf U}_1$ on the first register and ${\mathbf U}_2$ on
the second.

Consider that we perform the experiment
\begin{equation}\label{eq:expe}
z = {\mathcal M}_1 \circ {\mathbf U}_G \ket{0}\ket{\mathbf 0}
\end{equation}
where ${\mathcal M}_1$ denotes a measurement of the first register with
outcome~$z$ and where the first register initially holds the identity of
the group~$G$.

{\samepage{If~$G = \integer_2^n$ then the outcome $z \in G$ is a random 
element of
the orthogonal subgroup~$H_0^\perp \subgroup G$ by the discussion of
Simon's subroutine in Section~\ref{sec:alg}.  With the help of the
commutative laws in Theorem~\ref{thm:comm}, we now give a short proof
that this holds for every finite additive Abelian group~$G$.}}

The experiment given in Equations~\ref{eq:simongeneral}
and~\ref{eq:expe} consists of four operations.  As~the first, we apply
${\mathbf F}_G^{-1} \otimes {\mathbf I}$ on the initial zero state
\ket{0}\ket{\mathbf 0}, producing an equally-weighted superposition of
all elements in the group~$G$,
\[\sfrac{|G|} \sum_{g \in G} \ket{g}\ket{\mathbf 0}.\]
Then, as the second operation, applying ${\mathbf U}_\rho$ gives a
superposition of all cosets of~$H_0$,
\[\sfrac{|T_0|} \sum_{t \in T_0} \ket{t + H_0}\ket{\rho(t)}
  = \sfrac{|T_0|} \sum_{t \in T_0} (\tau_t \ket{H_0})\ket{\rho(t)}.\]
Here~$T_0$ denotes a transversal for~$H_0$ in~$G$.  Applying, as the
third operation, the Fourier transform on the first register produces
the superposition
\begin{align*} \ket{\Psi} 
&= \sfrac{|T_0|} \sum_{t \in T_0} ({\mathbf F}_G \circ \tau_t \ket{H_0})
  \ket{\rho(t)}\\
&= \sfrac{|T_0|} \sum_{t \in T_0} (\phi_t \circ {\mathbf F}_G \ket{H_0})
  \ket{\rho(t)}\\
&= \sfrac{|T_0|} \sum_{t \in T_0} (\phi_t \ket{H_0^\perp})
  \ket{\rho(t)}.
\end{align*}

Since the operator~$\phi_t$ changes only phases and not amplitudes, a
measurement of $\phi_t \ket{H_0^\perp}$ gives the same probability
distribution on the possible outcomes as a measurement of~$\ket{H_0^\perp}$.
It~follows that the outcome $z = {\mathcal M}_1
\ket{\Psi}$ is a random element of the orthogonal subgroup~$H_0^\perp$.
This completes our short proof of how the natural generalization of
Simon's subroutine can be used to sample random elements of~$H_0^\perp$.

The time needed to apply operator~${\mathbf U}_G$ is equal to twice
the time to compute~${\mathbf F}_G$ plus the time to compute the
function~$\rho$.  By~a result of Kitaev~\cite{Kitaev95}, for all finite
additive Abelian groups~$G$, the Fourier transform~${\mathbf F}_G$ can
be applied in polynomial time in~$\log|G|$.  However, his method
applies the transform not with perfection, but only with arbitrary good
precision (see~\cite{Kitaev95} for details).  Yet, this suffices to
imply a \ZQP--algorithm for the Abelian subgroup problem.

A~direct generalization of our \QP--algorithm would \mbox{require} the
solutions to two problems.

The first problem is that we must be capable of computing the Fourier
transform~${\mathbf F}_G$ exactly.  Cleve~\cite{Cleve94} and
Coppersmith~\cite{Copp94}, building on 
\pagebreak
the work of Shor~\cite{Shor94}, showed that it can be applied 
exactly in polynomial time whenever $G$ is of smooth
order. Here the order of a group~$G$ is {\em smooth\/} if all its prime
factors are at most~$\log^c |G|$ for some fixed 
\mbox{constant~$c$.  Thus,}
in~that case we can also apply~${\mathbf U}_G$ 
\mbox{efficiently} and exactly,
assuming we are given a polynomial-time (in~$\log |G|$) algorithm
to compute~$\rho$. 

The second problem is how to make certain that we find larger
and larger subgroups of~$H_0^\perp$ at each iteration until we eventually
have a generating set for~$H_0^\perp$.  Suppose we have previously found the
subset~$Y \subseteq H_0^\perp$ and now we measure some element~$z \in
H_0^\perp$.  For the group $G=\integer_2^n$, we~ensured
in~Section~\ref{sec:zero}, via Lemma~\ref{lm:AA}, that $z$ is the zero element
of~$G$ if $Y$ generates~$H_0^\perp$, and otherwise $z \in H_0^\perp \setminus
\langle Y \rangle$ is not generated by~$Y$.  Thus, in the latter case $Y
\cup \{z\}$ generates a subgroup strictly larger than the one generated
by~$Y$ itself.

Lemma~\ref{lm:AA} implies that if we can find a function~$\chi$ defined
on~$G$ such that $\chi$ equals~$1$ on exactly half the elements
of~$H_0^\perp$ and $\chi$ is~$0$ on the subgroup generated by~$Y \subset
H_0^\perp$ then we can ensure that~$z$ is nonzero.  We~can show that
this implication holds not only with the above fraction~$1/2$, but for
any fraction~$1/k$ where $k \leq \log^c|G|$ for some fixed
constant~$c$.  

We~are currently investigating for which groups of smooth order we can
find such a function~$\chi$ since this would solve the second problem.
If, in addition, there is an \mbox{efficient} algorithm to compute~$\chi$ then
this would imply a \mbox{\QP--algorithm} for the group under consideration.

As~our final example of generalizing our \QP--algorithm for Simon's
subgroup problem, consider the discrete logarithm problem defined as
follows.  For every prime~$p$, let $\integer_p^\star$ denote the
multiplicative cyclic group of the positive \mbox{integers} smaller than~$p$.
The {\em discrete logarithm problem\/} is given~$p$, a generator~$\zeta$
of~$\integer_p^\star$, and an element $a \in \integer_p^\star$, find
\mbox{$0 \leq r < p$} such that \mbox{$\zeta^r = a$} in
\mbox{$\integer_p^\star$}.

Shor gave in~\cite{Shor94} a \ZQP--algorithm for this problem.  In~our
language, his solution consists in a reduction to a problem equivalent to
a special case of the Abelian subgroup problem, followed by an algorithm
for that problem.  Let~$G = \integer_{p-1}^2$ and define function
$\rho: G \rightarrow \integer_p^\star$~by
\[\rho((g_1,g_2)) = \zeta^{g_1} a^{g_2}\]
for $g=(g_1,g_2) \in G$.  Let $H_0 \subgroup G$ be the cyclic subgroup
of order~$p-1$ generated by the element~$(r,-1) = (r,p-2)$.  Then~$\rho$
is constant and distinct on each coset of~$H_0$.  The \mbox{orthogonal}
subgroup~$H_0^\perp$ has also order~$p-1$ and is generated by~$(1,r)$.
It~is now easy to see that the discrete logarithm problem reduces to
finding the unique element~\mbox{$(g_1,g_2) \in H_0^\perp$}
for which \mbox{$g_1=1$}.
We~can show that if we are given a quantum algorithm to compute~${\mathbf
F}_{p-1}$ \mbox{exactly} then we can find that unique element in worst-case
polynomial time (in~$\log p$) on a quantum computer.  Here~${\mathbf
F}_{p-1}$ denotes the quantum Fourier transform for the cyclic
group~$\integer_{p-1}$.

{\samepage
\begin{theorem}[\QP--algorithm for Discrete Logarithms]
Let $p$ be a prime and $\zeta \in \integer_p^\star$ be a generator.  Then
given a quantum algorithm to compute~${\mathbf F}_{p-1}$ exactly, there
exists a \QP--algorithm that, for all $a \in \integer_p^\star$\,, finds
$0 \leq r \!< p-1$ such that $\zeta^r = a$ in~$\integer_p^\star$\,.
\end{theorem}
}

Let us end this paper by posing the open problem of finding a
\QP--algorithm for prime factorization.

\section*{Acknowledgments}
We~are grateful to Michel Boyer and Alain Tapp for helpful comments.
The second author thanks Joan Boyar for valuable discussions and for her
interest in this work.  This research was carried out while the second
author was at the Laboratoire d'informatique th\'eorique et quantique at
Universit\'e de Montr\'eal and he thanks the faculty and the students
for their hospitality.

\end{document}